\documentclass[%
preprint,
 amsmath,amssymb,
 aps,
]{revtex4-2}

\usepackage{graphicx}
\usepackage{dcolumn}
\usepackage{bm}

\begin{document}

\preprint{APS/123-QED}

\title{Anomalous Hall effect with plateaus observed in a magnetic Weyl semimetal NdAlGe at low temperatures}

\author{
Naoki Kikugawa$^{1*}$\email{KIKUGAWA.Naoki@nims.go.jp}, 
Shinya Uji$^{2}$, 
Taichi Terashima$^{2}$
}
 \affiliation{
$^{1}$Center for Basic Research on Materials (CBRM), National Institute for Materials Science, 3-13 Sakura, Tsukuba, Ibaraki 305-0003, Japan\\
$^{2}$Research Center for Materials Nanoarchitectonics (MANA), National Institute for Materials Science, 3-13 Sakura, Tsukuba, Ibaraki 305-0003, Japan
}



\date{\today}

\begin{abstract}
In the $R$Al(Si,Ge) ($R$: lanthanides) family, both spatial inversion and time-reversal symmetries 
are broken. 
This may offer opportunities to study Weyl-fermion physics in nontrivial spin structures emerging 
from a noncentrosymmetric crystal structure. 
In this study, we investigated the anomalous Hall effect (AHE) in NdAlGe via magnetotransport, 
magnetization, and magnetic torque measurements down to 40\,mK (0.4\,K for magnetization). 
The single crystals grown by a laser-heated floating-zone method exhibit 
a single magnetic phase transition at $T_{\rm M}$\,=\,13.5\,K, 
where the $T_{\rm M}$ is the transition temperature. 
With the magnetic field parallel to the easy $\lbrack$001$\rbrack$ axis, 
the AHE gradually evolves as the temperature decreases below $T_{\rm M}$. 
The anomalous Hall conductivity (AHC) reaches $\sim$320\,$\Omega^{-1}$\,cm$^{-1}$ at 40\,mK 
in the magnetically saturated state. 
Except in low-temperature low-field plateau phases, 
the AHC and magnetization are proportional, 
and their ratio agrees with the ratios for conventional ferromagnets, 
suggesting that the intrinsic AHE occurs by the Karplus-Luttinger mechanism. 
Below $\sim$0.6\,K, the curves of Hall resistivity against the field exhibit plateaus at low fields 
below $\sim$0.5\,T, 
correlating with the plateaus in the magnetization curve. 
For the first plateau, the magnetization is one order of magnitude smaller 
than the magnetically saturated state, 
whereas the AHE is more than half that in the saturated state. 
This finding under well below $T_{\rm M}$ suggests that 
the AHE at the first plateau is not governed by the magnetization and 
may be interpreted based on a multipole or spin chirality.

\end{abstract}

\keywords{magnetic Weyl semimetal, anomalous Hall effect, NdAlGe}
\maketitle


\section{Introduction}

Topologically nontrivial phases in condensed matter have attracted much attention recently 
due to their novel physical properties 
\cite{Wang_PRB_2013,Burkov_NatMater_2016,Muchler_AngewandteChemie_2012, 
Hasan_AnnualReview_2011,Ando_JPSJ_2013,Yan_AnnualReview_2017,Armitage_RevModPhys_2018}. 
Weyl semimetals are one such class of materials. 
These semimetals have band crossings near the Fermi level and host emergent 
relativistic quasiparticles, namely, Weyl fermions. 
The Weyl fermions can be realized when either spatial inversion or time-reversal symmetry is broken. 
Magnetic Weyl semimetals breaking time-reversal symmetry further offer opportunities 
to study the interplay between magnetic interactions and topologically nontrivial electronic structures; 
exhibiting novel phenomena such as the anomalous Hall and Nernst effects 
with no (or negligibly small) magnetization, 
optical Hall conductivity, and presence of axion insulators, and chiral domain walls 
\cite{Nagaosa_NatRevMater_2020,Bernevig_Nature_2022,Nakatsuji_AnnualReview_2022,Manna_NatRevMater_2018,Okamura_NatCommun_2020}. 
These phenomena may provide the basis for the next-generation spintronics applications 
\cite{Giustino_JPhysMater_2020,He_NatMater_2022,Smejkal_NatPhys_2018}. 

The $R$Al(Si,Ge) ($R$: lanthanides) family with the space group $I$4$_{1}md$ (No.\,109) is 
a new class of magnetic Weyl semimetals 
where both the inversion and time-reversal symmetries are broken 
\cite{Puphal_PhyRevMater_2019,Chang_PhysRevB_2018}. 
Recent studies have suggested a topological magnetic order in SmAlSi 
\cite{Yao_PhysRevX_2023}  and NdAlSi \cite{Gaudet_NatMater_2021,Wang_PhyRevB_2022}, 
topological Hall effect in SmAlSi \cite{Yao_PhysRevX_2023} 
and CeAl(Si,Ge) \cite{Yao_PhysRevX_2023,Piva_PhysRevMater_2023,Piva_PhyRevRes_2023}, 
anomalous Hall and Nernst effects in PrAl(Ge,Si) and NdAl(Si,Ge) 
\cite{Meng_APLMater_2019,Yang_APRMater_2020,Destraz_npjQuantumMater_2020}, 
anomalous thermal conductivity \cite{Tanwar_arXiv_2023}, 
unusual quantum oscillations in NdAlSi \cite{Wang_arXiv_2022,Zhang_PhysRevRes_2023}, 
possible axial gauge fields in PrAlGe \cite{Destraz_npjQuantumMater_2020}, 
domain wall chirality in CeAl(Si,Ge) \cite{Piva_PhyRevRes_2023,He_SciChina_2023}, 
surface Fermi arcs and bulk Weyl fermion dispersion in PrAlGe 
\cite{Sanchez_NatCommun_2020} and NdAlSi \cite{Li_arXiv_2023}, 
and reconstruction of the electronic structure across the magnetic transition in PrAlGe 
\cite{Yang_npjQuantumMater_2022}. 
We focused on NdAlGe in this study. 
The physical properties of NdAlGe were investigated by several groups 
using flux-grown crystals 
\cite{Zhao_NewJPhys_2022,Yang_PhysRevMater_2023,Cho_SSRN_2022,Dhital_PhyrevB_2023}. 
Yang \textit{et al}. \cite{Yang_PhysRevMater_2023} reported that NdAlGe undergoes 
two successive magnetic transitions: 
an incommensurate spin-density-wave transition at $T_{\rm ic}$\,=\,6.8\,K, 
where the spin structure is predominantly aligned in the (001) direction 
with small helical canting to the in-plane, 
and a commensurate ferrimagnetic transition at $T_{\rm com}$\,=\,5.1\,K, 
where the spin structure becomes a ``down-up-up'' structure propagating in $\lbrack$110$\rbrack$ 
or $\lbrack$1--10$\rbrack$ direction. 
The polarized ``up-up-up'' structure in high fields is formed through a metamagnetic-like behavior under a magnetization of $\sim$3\,T. 
Dhital \textit{et al}. reported slightly different transition temperatures $T_{\rm ic}$\,=\,6.3 K 
and $T_{\rm com}$\,=\,4.9\,K \cite{Dhital_PhyrevB_2023}. 
The anomalous Hall effect (AHE) was observed in both the ``down-up-up'' and polarized regions. 
This finding contrasts with the fact that the AHE was not detected in a sister material, NdAlSi 
\cite{Yang_PhysRevMater_2023}. 
Yang \textit{et al}. argued that 
the AHE in the ``down-up-up'' and polarized regions was governed 
by an intrinsic and extrinsic origin, respectively, 
whereas Dhital \textit{et al}. proposed an intrinsic origin of both regions 
\cite{Yang_PhysRevMater_2023,Dhital_PhyrevB_2023}. 
Two other research groups also grew NdAlGe crystals by a flux method, but they observed 
a single magnetic transition at 5.2 K \cite{Zhao_NewJPhys_2022} or 6 K \cite{Cho_SSRN_2022}. 
Thus, we see that flux-grown crystals exhibit considerable sample dependence. 
Recently, we have succeeded in growing the single crystals of NdAlGe by a laser-heated floating-zone technique \cite{Kikugawa_inorganics_2023}. 
This technique can minimize accidental contamination by impurities, allowing us to study the intrinsic properties of NdAlGe. 
We describe magnetotransport and magnetic torque measurements down to 40\,mK, 
and magnetization ones to 0.4\,K. 
We observed that AHE develops below a single magnetic transition temperature. 
We argue that, in all regions except a low-temperature low-field region, the observed AHE 
can be ascribed to the intrinsic Berry curvature and is consistent with the Karplus-Luttinger theory 
developed for ferromagnets \cite{Karplus_PhysRev_1954}. 
At low temperatures and under low fields, 
the Hall conductivity versus magnetic field curves exhibit plateaus. 
A large AHE is observed at the first plateau, despite low magnetization. 
We discuss the possible explanations for this phenomenon.  

\section{Experimental Details}

Single crystals of NdAlGe were grown by a laser-heated floating-zone method. 
The detailed growth procedure was described in \cite{Kikugawa_inorganics_2023}. 
Here, the grown crystals are deficient in aluminum due to the evaporation during growth. 
This is in sharp contrast to that aluminum-rich crystals were obtained by flux methods 
\cite{Zhao_NewJPhys_2022,Yang_PhysRevMater_2023,Cho_SSRN_2022,Dhital_PhyrevB_2023}. 
The crystals were cut and polished into rectangles with a typical size of 
3.4\,$\times$\,0.5\,$\times$\,0.4\,mm$^{3}$ in 
$\lbrack$100$\rbrack$, $\lbrack$010$\rbrack$], and $\lbrack$001$\rbrack$ directions, respectively. 
The electrical contacts were spot-welded and supported with silver paste. 
The contact resistances were below 0.1\,$\Omega$. 
The magnetoresistivity and Hall resistivity were measured simultaneously using a low-frequency 
($\sim$13\,Hz) AC method. 
The measurements were performed in a top-loading $^{3}$He\,-\,$^{4}$He dilution refrigerator 
at temperatures ($T$) down to 40\,mK with sweeping magnetic field ($H$) between $-$17.5 
and $+$17.5\,T. 
The electrical current and magnetic field were applied in the $\lbrack$100$\rbrack$ 
and $\lbrack$001$\rbrack$ directions, respectively, unless specified otherwise. 
Because the misalignments of the contact electrodes can cause mixing of the magnetoresistivity and Hall resistivity, the magnetoresistivity ($\rho_{xx}$) and Hall resistivity ($\rho_{yx}$) were obtained by 
symmetrizing $\rho_{xx}$\,=\,(($\rho_{xx}^{\rm exp}(H)$\,$+$\,$\rho_{xx}^{\rm exp}(-H)$)/2 and 
antisymmetrizing the experimental data 
$\rho_{yx}$\,=\,(($\rho_{yx}^{\rm exp}(H)$\,$-$\,$\rho_{yx}^{\rm exp}(-H)$)/2, respectively. 
The magnetic torque ($\tau$) was measured using a capacitive cantilever method 
in a top-loading $^{3}$He\,-\,$^{4}$He dilution refrigerator. 
Because the magnetic torque vanishes for symmetric directions, the measurements were performed 
under the magnetic field applied 3$^{\circ}$ off from the exact $\lbrack$001$\rbrack$ 
to $\lbrack$010$\rbrack$ direction. 
The isothermal magnetization ($M$) measurements under 
$H$\,$\parallel$\,$\lbrack$001$\rbrack$ between $-$16 and $+$16 T were performed down to 2\,K 
using the options of Physical Property Measurement System (PPMS, Quantum Design). 
The measurement at 0.4\,K was performed using the Magnetic Property Measurement System 
(MPMS3, Quantum Design) with a $^{3}$He cooling option. 
The specific-heat ($C_{P}$) under zero field was measured down to 0.4\,K 
using the options of Physical Property Measurement System (Quantum Design). 
The measurement was performed by a relaxation method. 

\section{Results}

Figure 1(a) shows the temperature ($T$) dependence of resistivity $\rho_{xx}$ under zero field. 
A clear kink observed at 13.5\,K corresponds to the magnetic ordering temperature ($T_{\rm M}$). 
The inset of Fig. 1(a) shows a closeup of $\rho_{xx}$ below 2\,K, 
where no anomalies are seen down to 40\,mK. 
This finding is consistent with the specific-heat measurement of our floating-zone NdAlGe crystal 
down to 0.4\,K (see Fig. S1 in the Supplemental Material \cite{SupplementalMaterial}), 
revealing a single sharp transition at $T_{\rm M}$. 
Notably, the transition width of the specific-heat jump at $T_{\rm M}$ is as sharp as 0.4\,K, 
suggesting that the floating-zone crystal in this study is highly homogeneous. 
In comparison, two successive transitions were observed at lower temperatures ($\sim$5\,K and 
$\sim$6\,$-$\,7\,K) 
\cite{Yang_PhysRevMater_2023,Dhital_PhyrevB_2023}, 
or a single transition was observed at 5\,$-$\,6 K in flux-grown NdAlGe crystals 
\cite{Zhao_NewJPhys_2022,Cho_SSRN_2022}. 
Figure 1(b) shows the Hall resistivity $\rho_{yx}$ at 40\,mK in a wide magnetic field ($H$) 
range of $-$17.5 to $+$17.5\,T.  
Hysteresis appears at low fields ($\textless$\,0.5\,T), which will be detailed below. 
At $\lvert \mu_{0}H \rvert$\,$\textgreater$\,1\,T ($\mu_{0}$: magnetic permeability in vacuum), 
the $\rho_{yx}$ exhibits a linear field dependence with a positive slope, 
suggesting that holes are the dominant carrier. 
The ordinary Hall coefficient ($R_{0}$) deduced under high field greater than 1\,T is 
$+$1.28\,$\times$\,10$^{-3}$\,cm$^{3}$/C. A similar value of $+$1.25\,$\times$\,10$^{-3}$\,cm$^{3}$/C was obtained 
for another sample (Fig. S2(a) \cite{SupplementalMaterial}). 
The coefficients correspond to a carrier density of $+$4.88 to $+$4.99\,$\times$\,10$^{21}$\,cm$^{-3}$, 
assuming a single band. In comparison, 
larger values of $R_{0}$ ranging from $+$4 to $+$7\,$\times$\,10$^{-3}$\,cm$^{3}$/C, 
corresponding to smaller hole densities, 
were reported for flux-grown crystals 
\cite{Yang_PhysRevMater_2023,Cho_SSRN_2022,Dhital_PhyrevB_2023}. 
The difference of the values is likely related to the fact that while floating-zone crystals are 
aluminum deficient \cite{Kikugawa_inorganics_2023}, 
flux-grown ones are aluminum rich 
\cite{Zhao_NewJPhys_2022,Yang_PhysRevMater_2023,Cho_SSRN_2022,Dhital_PhyrevB_2023}. 
No anomaly was found in $\rho_{yx}$ for fields above 1\,T (Fig. 1(b)); 
this result is consistent with the magnetization ($M$) curve of our floating-zone crystal (Fig. S3 \cite{SupplementalMaterial}) 
and is in sharp contrast with the fact that $\rho_{yx}$ and $M$ in flux-grown crystals exhibit 
an anomaly around 3\,T 
\cite{Zhao_NewJPhys_2022,Yang_PhysRevMater_2023,Cho_SSRN_2022,Dhital_PhyrevB_2023}.  
Figure 2(a) shows the Hall resistivity $\rho_{yx}$ in a low-field region at selected temperatures 
between 1.5\,K and 20\,K. 
The anomalous Hall contribution is observed below $T_{\rm M}$\,=\,13.5\,K. 
The hysteresis between up and down field-sweep is obvious below 10\,K and 
is closely linked to the magnetization hysteresis (Fig. S3(a) \cite{SupplementalMaterial}). 
We notice the $\rho_{yx}$ exhibits nonlinear behavior in the hysteretic region at 1.5 and 5\,K. 
Figure 2(b) shows the $\rho_{yx}$ curves below 1.5\,K. 
Clear plateaus are developed in the hysteretic region between $-$0.5 and $+$0.5\,T 
as the temperature decreases. 
As the field is swept from the negative to positive values at the lowest temperature of 40\,mK, 
the $\rho_{yx}$ jumps to the first plateau ($\rho_{yx}$\,=\,1.1\,$\mu\Omega$\,cm), 
second ($\rho_{yx}$\,=\,1.8\,$\mu\Omega$\,cm), and third ($\rho_{yx}$\,=\,2.5\,$\mu\Omega$\,cm) plateaus 
at 0.07\,T, 0.22\,T, and 0.36\,T, respectively. 
Finally, $\rho_{yx}$ reaches 2.7\,$\mu\Omega$\,cm in the field-induced polarized state above 0.5\,T. 
The plateau behavior in $\rho_{yx}$ below 1\,K was reproduced in another sample, 
and the $\rho_{yx}$ finally reached 2.7\,$\mu\Omega$\,cm above 0.5\,T (Fig. S2(b) \cite{SupplementalMaterial}). 
Such plateau behavior has not been reported in the flux-grown crystals. 
Figure 2(c) shows 
the Hall conductivity $\sigma_{xy}$ at 40\,mK, 
calculated using the formula $\sigma_{xy}$\,=\,$\rho_{yx}/(\rho_{xx}^2 +\rho_{yx}^2)$, 
and the inset shows the behavior of magnetoconductivity 
$\sigma_{xx}$\,=\,$\rho_{xx}/(\rho_{xx}^2 +\rho_{yx}^2)$. 
The Hall conductivity $\sigma_{xy}$ reaches 
a large value of 320\,$\Omega^{-1}$\,cm$^{-1}$ in the high-field polarized state. 
For the polarized state of flux-grown crystals, $\sigma_{xy}$ is 
$\sim$1,030 $\Omega^{-1}$\,cm$^{-1}$ \cite{Dhital_PhyrevB_2023} or 
0.5\,$-$\,1.5\,$\times$\,10$^{3}$\,$\Omega^{-1}$\,cm$^{-1}$ \cite{Yang_PhysRevMater_2023}. 
No anomaly was seen in the magnetoconductivity $\sigma_{xx}$ in the same field region 
(inset of Fig. 2(c)). 
Figure 3 shows the magnetic field angle dependence of $\rho_{yx}$ at a base temperature of 40\,mK. 
Herein, the field angle ($\theta$) was measured 
from $\lbrack$001$\rbrack$ to $\lbrack$010$\rbrack$ (inset of Fig. 3). 
The curves measured at different angles collapse into a single curve when plotted against 
$\mu_{0}$$H$cos$\theta$. 
This finding is consistent with the Ising character of the neodymium magnetism in NdAlGe 
\cite{Zhao_NewJPhys_2022,Yang_PhysRevMater_2023,Cho_SSRN_2022,Dhital_PhyrevB_2023,Kikugawa_inorganics_2023}. 
Figure 4 shows the plot of anomalous Hall resistivity ($\rho_{yx}^{\rm AHE}$) against temperature. 
Herein, the $\rho_{yx}^{\rm AHE}$ is defined by the extrapolations of the Hall resistivity 
from the high to the zero magnetic field for high temperatures, as presented in Fig. 2(a), 
and it is the residual Hall resistivity at $\mu_{0}H$\,=\,0\,T for low temperatures (Fig. 2(b)). 
The $\rho_{yx}^{\rm AHE}$ gradually evolves below $T_{\rm M}$ and reaches 2.7\,$\mu\Omega$\,cm 
at the lowest temperature. 
The inset of Fig. 4 shows $\sigma_{xy}^{\rm AHE}$ plotted against extrapolated zero-field magnetization 
$M_{0}$ for four temperatures ($T$\,=\,2, 10, 12, and 13\,K), where $M_{0}$ is obtained by 
extrapolating the high-field part of a magnetization curve toward zero field as presented in Fig. S3(a) \cite{SupplementalMaterial}. 
A clear linear relation exists between $\sigma_{xy}^{\rm AHE}$ and $M_{0}$. 

Figures 5(a)-5(c) show a comparison of the magnetization $M$, magnetic torque ($\tau$) 
divided by magnetic field $\tau$/($\mu_{0}H$), and the Hall resistivity $\rho_{yx}$ measured 
at $\sim$0.4\,K, the lowest possible temperature for magnetization measurements. 
The magnetic torque is given by $\vec{\tau} = \vec{M} \times \vec{H}$. 
Considering the Ising nature of the neodymium moments 
\cite{Zhao_NewJPhys_2022,Yang_PhysRevMater_2023,Cho_SSRN_2022,Dhital_PhyrevB_2023,Kikugawa_inorganics_2023}, 
$\tau$/($\mu_{0}H$) can be roughly approximated to $M$. 
Similar to the behavior of $\rho_{yx}$, 
as the field is increased from negative to positive values, both $M$ and $\tau$/($\mu_{0}H$) exhibit 
the first and second plateaus before the field-polarized state is reached above 0.5\,T. 
Although the $M$ and $\tau$/($\mu_{0}H$) values at the first plateau are close to zero, 
the $\rho_{yx}$ is more than half of the polarized state ($\sim$1.5\,$\mu$$\Omega$\,cm). 
The transition fields between the three regions slightly differ among the three measurements. 
This difference is mostly attributed to the sample difference and/or demagnetization factor differences. 
In addition, only the magnetization curve shows a rather wide transitional (nonflat) region 
from $\sim$0.18 to $\sim$0.28\,T between the first and second plateaus. 
This wide transition region was possibly caused by a temperature instability or increased temperature 
during the magnetization measurement; the samples were immersed in liquid $^{3}$He\,-\,$^{4}$He 
mixture for the magnetic torque and $\rho_{yx}$ measurements 
using the top-loading dilution refrigerator, 
whereas a sample was held in low-pressure $^{3}$He gas atmosphere 
for the magnetization measurement, 
and thermal contact was provided by copper wires attached to the sample. 
Consequently, the temperature stability and accuracy may be worse 
in the magnetization measurements. 
Figure 6 shows the $\tau$/($\mu_{0}H$) curves measured at various temperatures below 1.1 K. 
Similar to the $\rho_{yx}$ data (Figs. 2(b) and S2(b) \cite{SupplementalMaterial}), 
the plateaus are sensitive to temperature: 
the first and second plateaus survive only up to 0.5 and 0.4\,K, respectively. 
The heights of the plateau vary slightly with temperature.  

\section{Discussion}

We first argue that the observed anomalous Hall conductivity (AHC) 
$\sigma_{xy}^{\rm AHE}$\,$\sim$\,320\,$\Omega^{-1}$\,cm$^{-1}$ in the polarized state (Fig. 2(c)) 
is primarily attributed to the intrinsic Berry curvature. 
A theoretical scaling relation between $\sigma_{xy}$ and $\sigma_{xx}$ 
\cite{Onoda_PhyRevB_2008,Nagaosa_RevModPhys_2010,Chen_NatCommun_2021}, 
experimentally supported, indicated three regimes. 
AHE in each regime is dominated by different mechanisms: 
in the high-conductivity regime ($\sigma_{xx}$\,$\textgreater$\,10$^{6}$\,$\Omega^{-1}$\,cm$^{-1}$), 
skew scattering dominates AHE and $\sigma_{xy}^{\rm AHE}$\,$\propto$\,$\sigma_{xx}$; 
in the good-metal regime ($\sigma_{xx}$ is 10$^{4}$\,$-$\,10$^{6}$\,$\Omega^{-1}$\,cm$^{-1}$) 
where the intrinsic Berry-phase contribution dominates and 
$\sigma_{xy}^{\rm AHE}$ is approximately independent of $\sigma_{xx}$; 
in the bad metal regime ($\sigma_{xx}$\,$\textless$\,10$^{4}$\,$\Omega^{-1}$\,cm$^{-1}$) 
where $\sigma_{xy}^{\rm AHE}$\,$\propto$\,$\sigma_{xx}^{1.6}$. 
The AHC in the good-metal regime is of the order of 
10$^{2}$\,$-$\,10$^{3}$\,$\Omega^{-1}$\,cm$^{-1}$. 
Our results ($\sigma_{xx}$\,$\sim$\,1.1\,$\times$\,10$^{4}$$\Omega^{-1}$\,cm$^{-1}$ 
and $\sigma_{xy}^{\rm AHE}$\,$\sim$\,320\,$\Omega^{-1}$\,cm$^{-1}$ at $T$\,=\,40\,mK) 
perfectly fit the good-metal regime. 
The intrinsic Berry curvature contribution to the AHE in the polarized state of NdAlGe was theoretically calculated as $\sigma_{xy}^{\rm AHE}$\,$\sim$\,200\,$\Omega^{-1}$\,cm$^{-1}$ 
\cite{Yang_PhysRevMater_2023} or $\sim$\,270\,$\Omega^{-1}$cm$^{-1}$ 
\cite{Dhital_PhyrevB_2023}. 
These values are in good agreement with our experimental values. 
Furthermore, according to the Karplus-Luttinger theory of the intrinsic AHE, 
$\rho_{yx}^{\rm AHE}$\,$\sim$\,$\rho_{xx}^{2}$$M$ \cite{Karplus_PhysRev_1954}. 
Considering $\rho_{xx}$\,$\gg$\,$\lvert$$\rho_{yx}$$\rvert$, 
this relation indicates that $\sigma_{xy}^{\rm AHE}$ 
(approximately equal to $\rho_{yx}^{\rm AHE}/\rho_{xx}^{\rm 2}$) is proportional to $M$. 
The present data nicely satisfy this relation (inset of Fig. 4), 
further supporting the intrinsic mechanism. 
For various conventional ferromagnets exhibiting AHE, 
the ratio $\sigma_{xy}^{\rm AHE}/M$ is in the range of a few tens to 
1\,$\times$\,10$^{3}$\,$\Omega^{-1}$\,cm$^{-1}/$T (where $M$ is measured in tesla) 
\cite{Nakatsuji_AnnualReview_2022}. 
For the polarized state of NdAlGe, 
where $M$\,=\,0.45\,T, the ratio is 7.1\,$\times$\,10$^{2}$\,$\Omega^{-1}$\,cm$^{-1}/$T, 
lying in this empirical range. 
Our conclusion regarding the intrinsic AHE agrees with the conclusion drawn 
from studies on flux-grown crystals by Dhital \textit{et al} \cite{Dhital_PhyrevB_2023}. 
Conversely, Yang \textit{et al}. \cite{Yang_PhysRevMater_2023} observed that the temperature dependence of the AHC 
in the polarized state of flux-grown crystals differs with the sample and does not follow 
the magnetization; 
they argued that AHE in the polarized state of NdAlGe has an extrinsic origin. 
Next, we focus on the first plateau in the Hall resistivity. Figure 5 indicates that $\tau/(\mu_{0}H)$ 
and $M$ of the first plateau are close to zero and the magnetization is negative, respetively. 
Nevertheless, the $\rho_{yx}$ has a substantial value of $\rho_{yx}$\,$\sim$\,1.5\,$\mu\Omega$\,cm, 
$\sim$55\% of its value in the polarized state and has the same positive sign. 
For the first plateau, where $M$\,=\,0.043\,T, 
the ratio $\sigma_{xy}^{\rm AHE}/M$ is 4.3\,$\times$\,10$^{3}$\,$\Omega^{-1}$\,cm$^{-1}/$T, 
exceeding the aforementioned empirical range. 
Recent studies on anomalous Hall antiferromagnets indicated that the contribution 
of the Berry curvature to AHE is not necessarily controlled by magnetization 
\cite{Nakatsuji_AnnualReview_2022,Suzuki_PhysRevB_2017,Smejkal_NatRevMater_2022}. 
Mn$_{3}$Sn, for example, is an antiferromagnet without net magnetization 
(in the absence of spin-orbit coupling), but it exhibits a large AHE. 
In the case of such anomalous Hall antiferromagnets, 
breaking of the time-reversal symmetry that is necessary for AHE and is attributed to 
the magnetic multipoles composed of multiple atoms 
\cite{Suzuki_PhysRevB_2017,Smejkal_NatRevMater_2022}. 
Another example is CoTa$_{3}$S$_{6}$ and CoNb$_{3}$S$_{6}$ 
\cite{Ghimire_NatCommun_2018,Takagi_NatPhys_2023}. 
Arguably, a fictitious magnetic field associated with a scalar spin chirality induces a giant Hall response 
under the small net magnetization \cite{Takagi_NatPhys_2023}. 
For flux-grown NdAlGe samples, 
neutron diffraction measurements were performed, 
and the low-temperature zero-field spin structure was found to be composed 
of basically antiferromagnetic spin ``down-up-up'' chains along $\lbrack$110$\rbrack$ directions. 
Due to the Dzyaloshinskii-Moriya interaction, a part of the spins exhibits a slight canting 
from the $\lbrack$001$\rbrack$ direction, 
resulting in a noncollinear spin structure \cite{Yang_PhysRevMater_2023,Dhital_PhyrevB_2023}. 
The spin structure at the first plateau of the floating-zone crystals may be a variant 
with the same ``down-up-up'' motif. 
Determining the exact spin structure to identify the multipole or spin chirality responsible for 
the AHE of the first plateau is highly desirable. 
Finally, we discuss the differences between floating-zone and flux-grown crystals. 
The floating-zone crystals are aluminum deficient, 
whereas the flux-grown ones are aluminum-rich. 
This difference likely explains the smaller ordinary Hall coefficient $R_{0}$ (larger hole density) 
in the floating-zone crystals, and may also account for the smaller $\sigma_{xy}^{\rm AHE}$ 
in the polarized state of these floating-zone crystals 
since the theoretical calculations indicate that the AHC caused 
by Berry curvature decreases with a decrease in the Fermi level 
\cite{Yang_PhysRevMater_2023,Dhital_PhyrevB_2023}. 
Further, because the Ruderman-Kittel-Kasuya-Yosida (RKKY) interaction 
mediated by itinerant carriers is reasonably assumed to be the primary exchange interaction 
between neodymium moments, 
the differences between the magnetic phase diagrams of the floating-zone and flux-grown crystals 
can be attributed to their different carrier densities.  

\section{Summary}

we performed magnetotransport and magnetic torque measurements down to 40\,mK 
and magnetization ones to 0.4\,K on floating-zone single crystals of NdAlGe. 
We observed only one magnetic phase transition at $T_{\rm M}$\,=\,13.5\,K, 
in contrast to the two transitions observed in some of the flux-grown crystals 
\cite{Yang_PhysRevMater_2023,Dhital_PhyrevB_2023}. 
The AHE occurred below $T_{\rm M}$, and the AHC reached $\sim$\,320\,$\Omega^{-1}$\,cm$^{-1}$ 
at 40\,mK in the polarized state, 
comparable to the \textit{ab-initio} calculations of the intrinsic Berry curvature contribution 
\cite{Yang_PhysRevMater_2023,Dhital_PhyrevB_2023}. 
A comparison with the theoretical scaling relation between $\sigma_{xy}^{\rm AHE}$ 
and $\sigma_{xx}$ supports the Berry curvature origin of the AHE. 
A linear relation exists between $\sigma_{xy}$ and $M_{0}$, and their ratio 
$\sigma_{xy}^{\rm AHE}/M$ is in the typical range for ferromagnets. 
These results indicate that the AHE, except for the low-temperature low-field region, 
occurs within the framework of the Karplus-Luttinger theory. 
At low temperatures and fields below $\sim$0.6\,K and $\sim$0.5\,T, 
we observe plateaus in the curves of the Hall resistivity against the field. 
These plateaus in the Hall resistivity are correlated with the ones in the magnetization curves. 
In the first plateau, although the magnetization is one order of magnitude smaller than that 
in the saturated state, 
we observe a large anomalous Hall resistivity, 
more than half the value observed in the magnetically saturated state. 
This imbalance between the anomalous Hall effect and magnetization is analogous 
to anomalous Hall antiferromagnets such as Mn$_{3}$Sn or Co(Nb,Ta)$_{3}$S$_{6}$. 
This finding indicates that a multipole or spin chirality governs the AHC in the first plateau, 
and theoretical work on this topic is necessary in the future. 

\begin{acknowledgments}

We acknowledge Takanobu Hiroto for comments and support, Masao Arai, and Jun-ichi Inoue 
for comments and discussion, Ayumi Kawaguchi, Takashi Kato, Momoko Hayashi, 
Hitoshi Yamaguchi, Takeshi Shimada, Akira Kamimura, John McArthur, and Noritaka Kimura for support. 
This work is funded by KAKENHI Grants-in-Aids for Scientific Research 
(Grant Nos. 18K04715, 21H01033, and 22K19093), 
and Core-to-Core Program (No. JPJSCCA20170002) 
from the Japan Society for the Promotion of Science (JSPS) and 
by a JST-Mirai Program (Grant No. JPMJMI18A3). 
MANA was established by World Premier International Research Center Initiative (WPI), MEXT, Japan. 

\end{acknowledgments}

\bibliography{NdAlGe_Main_Reference_Revised}

\clearpage

\begin{figure}
\includegraphics[width=70mm]{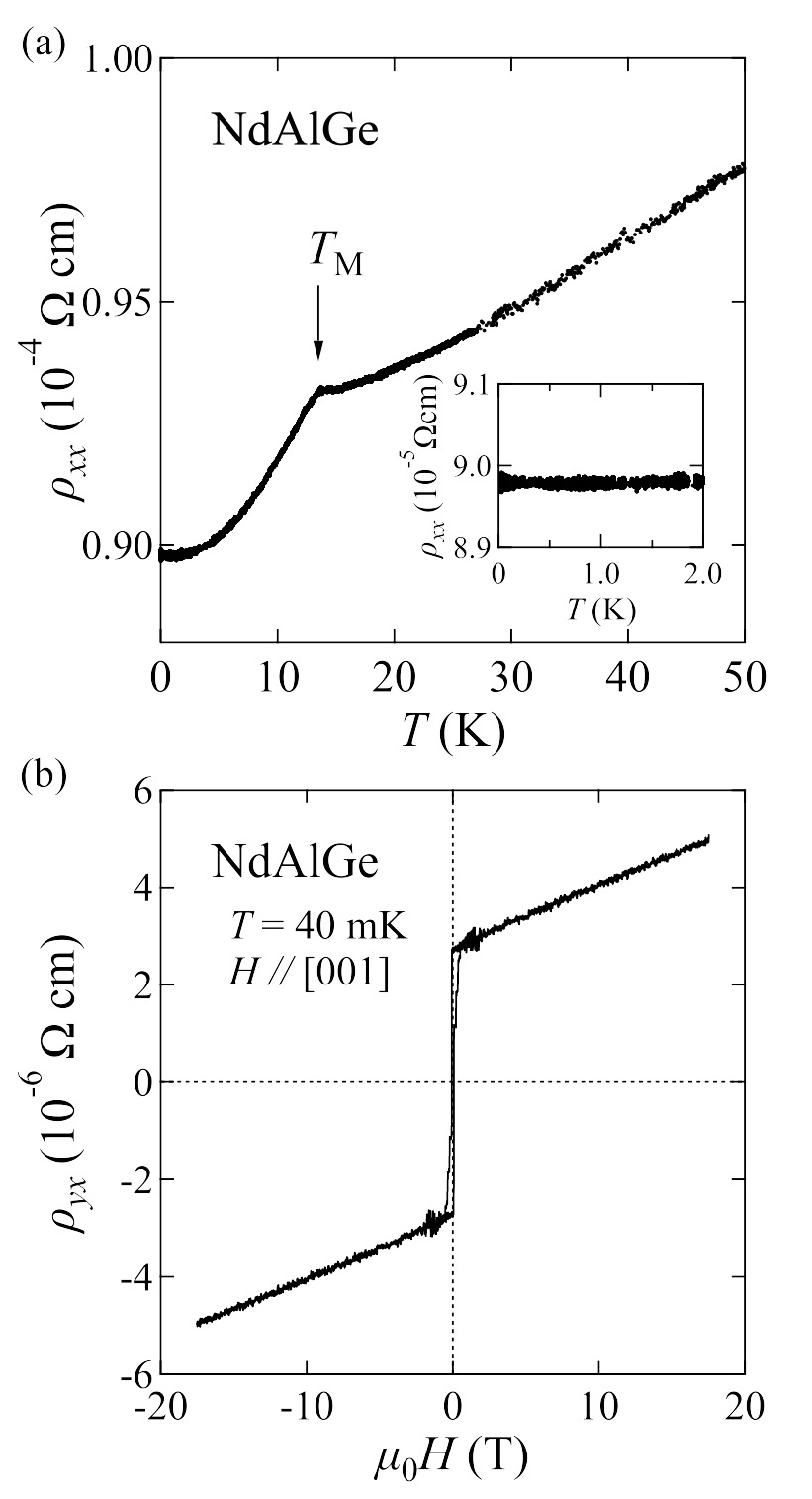}
\caption{
(a) Temperature dependence of resistivity $\rho_{xx}$ in NdAlGe under zero magnetic field; 
the current flows in the $\lbrack$100$\rbrack$ direction; inset: closeup of the resistivity 
between 40 mK and 2 K. 
(b) Hall resistivity $\rho_{yx}$ of NdAlGe at 40\,mK under the magnetic field 
between $-$17.5 and $+$17.5 T. 
The ordinary Hall coefficient $R_{0}$\,=\,$+$1.28\,$\times$\,10$^{-3}$\,cm$^{3}$/C 
is evaluated from the slope of the $\rho_{yx}$ above 1\,T. 
}
\end{figure}

\begin{figure*}
\includegraphics[width=160mm]{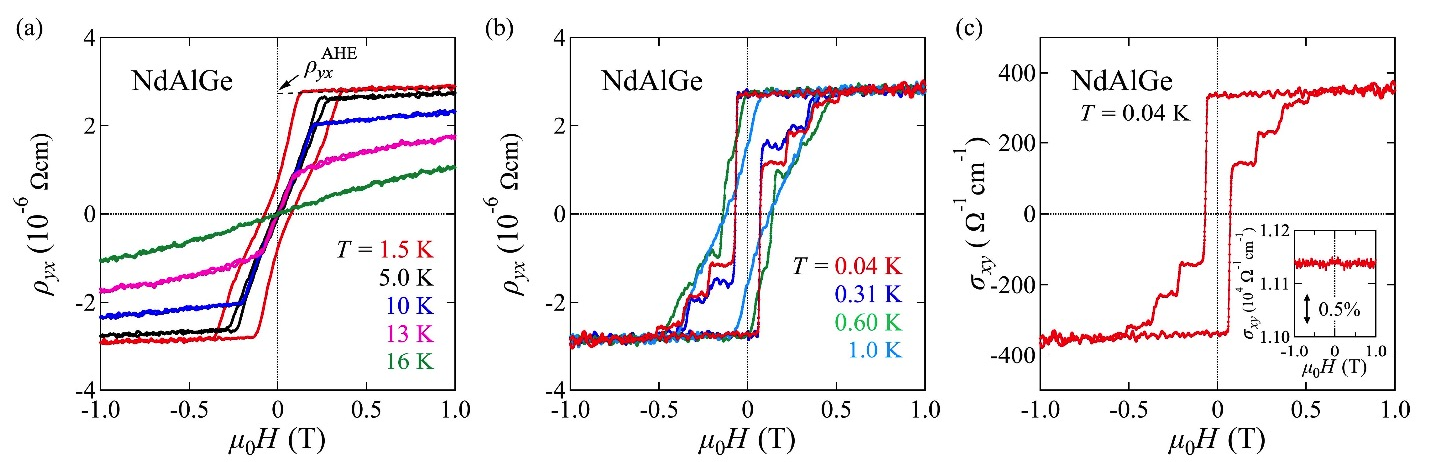}
\caption{
Hall resistivity $\rho_{yx}$ of NdAlGe at several temperatures 
(a) between 1.5 and 20\,K across the magnetic ordering temperature $T_{\rm M}$\,=\,13.5\,K, 
and (b) below 1\,K. 
the magnetic fields are applied along the $\lbrack$001$\rbrack$ direction; 
the anomalous Hall resistivity $\rho_{yx}^{\rm AHE}$ at high temperatures is defined 
by the extrapolations of the Hall resistivity from a high to zero magnetic field, 
as shown in (a). 
(c) Hall conductivity $\sigma_{xy}$\,=\,$\rho_{yx}/(\rho_{xx}^2 +\rho_{yx}^2)$ of NdAlGe at 40\,mK; 
inset presents the magnetoconductivity $\sigma_{xx}$\,=\,$\rho_{xx}/(\rho_{xx}^2 +\rho_{yx}^2)$.
}
\end{figure*}

\begin{figure}
\includegraphics[width=70mm]{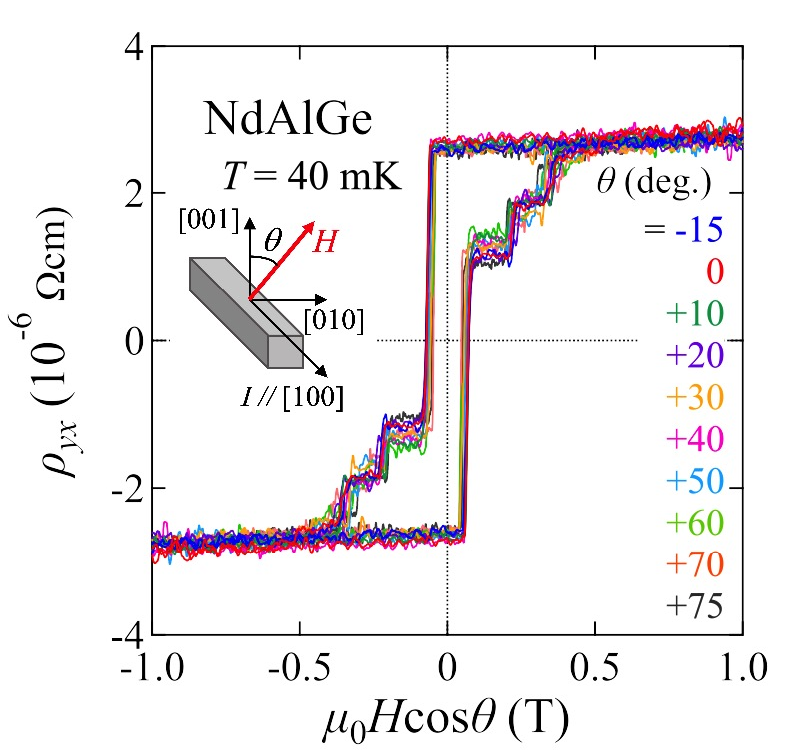}
\caption{
Magnetic field angle dependence of the Hall resistivity $\rho_{yx}$ in NdAlGe at 40\,mK. 
The field is tilted from the $\lbrack$001$\rbrack$ to $\lbrack$010$\rbrack$ axes as shown in the inset; 
the horizontal axis is $\mu_{\rm 0}H$cos$\theta$, where the $\theta$ is the tilting angle.
}
\end{figure}

\begin{figure}
\includegraphics[width=70mm]{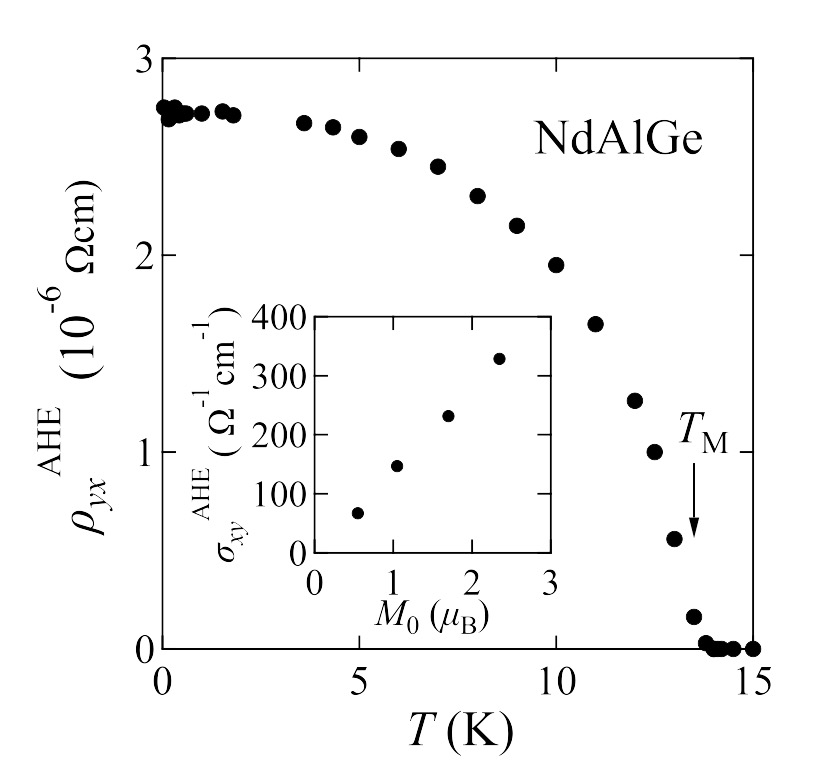}
\caption{
The anomalous Hall resistivity $\rho_{yx}^{\rm AHE}$ plotted against temperature; 
the $\rho_{yx}^{\rm AHE}$ gradually evolves below $T_{\rm M}$; 
inset: 
the anomalous Hall conductivity $\sigma_{xy}^{\rm AHE}$ against 
extrapolated zero-field magnetization $M_{0}$, 
obtained by extrapolating the high-field part of a magnetization curve toward zero field (Fig. S3(a)). 
}
\end{figure}

\begin{figure}
\includegraphics[width=70mm]{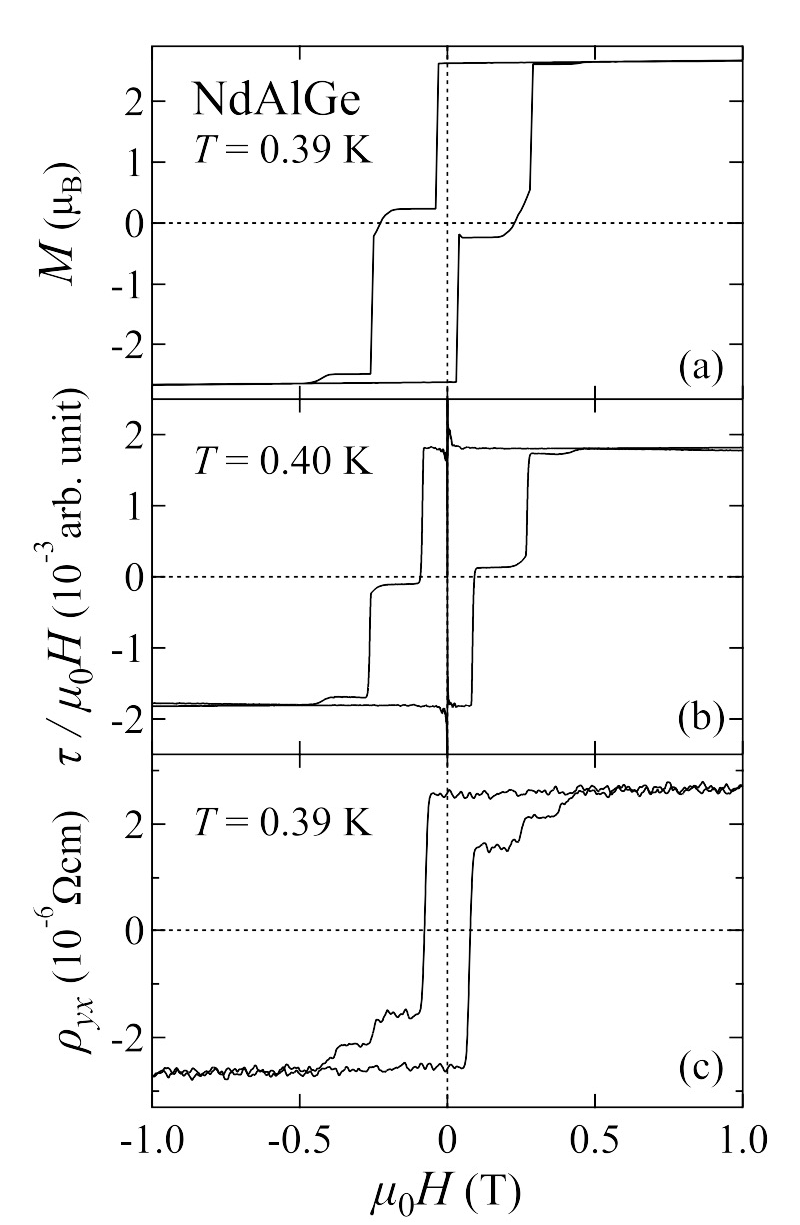}
\caption{
(a) Isothermal DC magnetization $M$, (b) magnetic torque ($\tau$) divided by field $\tau /\mu_{0}H$, 
and (c) Hall resistivity $\rho_{yx}$ of NdAlGe at $T$\,$\sim$\,0.4\,K. 
the plateau structures in the Hall resistivity are closely related to the magnetic properties. 
}
\end{figure}

\begin{figure}
\includegraphics[width=70mm]{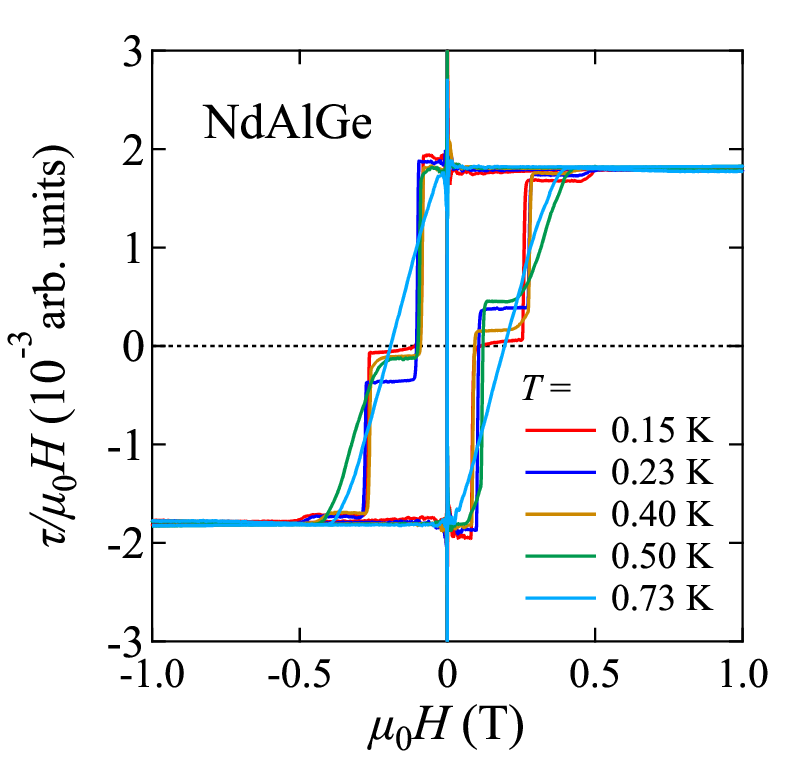}
\caption{
Magnetic torque divided by the applied field ($\tau /(\mu_{\rm 0}H)$) below 1 K. 
}
\end{figure}

\end{document}